# Enabling Multi-hop Forwarding in 6LoWPANs through Software-Defined Networking

Giacomo Tanganelli
Dipartimento di Ingegneria
dell'Informazione
University of Pisa
Pisa, Italy
giacomo.tanganelli@iet.unipi.it

Antonio Virdis
Dipartimento di Ingegneria
dell'Informazione
University of Pisa
Pisa, Italy
antonio.virdis@unipi.it

Enzo Mingozzi
Dipartimento di Ingegneria
dell'Informazione
University of Pisa
Pisa, Italy
enzo.mingozzi@unipi.it

*Abstract*—Wireless Sensor Networks (WSNs) play a major role in the expansion of the Internet of Things (IoT) market as they allow the deployment of low cost and low power networks possibly in large areas. The introduction of the 6LoWPAN standard allowed WSN to move into the IPv6 era, with a seamless integration of sensors into external networks. The recent application of IoT in industrial systems is now introducing new traffic patterns - machine to machine (M2M) above the others - and communication challenges that can modify the classical multipoint-to-point pattern of WSN, having the sink as the root of every communication. To solve these problems, we propose an architecture to enable an efficient and controlled multi-hop forwarding over 6LoWPANs. We base our approach on the standard 6LoWPAN protocol stack and we exploit the Software Defined Networking paradigm to achieve network reconfigurability. We evaluate the performance of our architecture in a simulated environment, with both multipoint-to-point and M2M traffic, and we demonstrate its feasibility in a real testbed.

## I. INTRODUCTION

The development of wireless sensor networks (WSNs) has exponentially increased in the last years and this trend continues today as the IoT world gains more and more attention, both in the commercial market and in the industrial one, with novel application scenarios ranging from smart homes, to warehouse management, to factory monitoring. Each new scenario comes with its own peculiarities in terms of deployment, traffic characteristics and requirements. Originally, typical communication patterns were multipoint-to-point, i.e. from sensor nodes to the sink, and point-to-multipoint, i.e. from the sink to sensor nodes [1]. Nowadays, instead, the application of WSN to industrial systems has introduced novel interaction pattern. As an example, machine-to-machine (M2M) communications, i.e. point-to-point, are logically contained within the WSN, without directly involving the sink. These new and specific scenarios forces manufacturers to create ad-hoc system for each application, optimizing network configuration to meet user's needs.

To solve the mentioned interoperability issue, the IETF has already standardized a suite of solutions that allow WSNs to seamlessly transport IP data packets. Specifically, the 6LoWPAN standard [2] has been defined as an adaptation layer to enable IPv6 end-to-end communications between external application and sensor nodes. Furthermore, the IETF has also defined the RPL routing protocol [3] as a dedicated routing protocol for low power and lossy networks. Through the adoption of this suite of protocols, WSNs - which are commonly referred to as 6LoWPANs - can be exploited without the need to deploy dedicated gateways for protocol translation, thus reducing interoperability issues.

On the other hand, a new paradigm of network softwarization has recently established as an innovative way to customize and manage network deployment and configuration via software in a centralized manner. This approach is called Software Defined Networking (SDN) [4] and is being extensively used in wired networks, e.g. in data centers, within core-networks of communication service providers, etc.. SDN gives network operators the ability to control the system in a centralized manner, having a *Controller* responsible of configuring network equipment, called SDN Switches, via dedicated control mechanism, thus removing the burden of complex distributed protocols for configuration. This does not only reduce the operational cost of the network, but also allows operator to modify the network configuration on the fly, responding quickly to user's need.

In the last years, the SDN technology has also been considered in the WSN domain, trying to apply the same softwarization approach to wireless networks. The SDN solutions that are commonly used in the wired context, e.g. OpenFlow [5], are not suitable for wireless networks [6], due to the non-negligible overhead associated with the control plane. For this reason, most of the available solutions [7][8] provide ad-hoc solutions for SDN in WSN by defining their own protocol stack, comprising custom realization for control plane and forwarding operations. The use of custom solutions, although aiming at reducing the overhead, limits significantly the interoperability of the system with the existing deployment and thus increases significantly the time-to-market.

In this work, we propose Software-Defined 6LoWPAN (SD-6LoWPAN), an architecture for integrating SDN into 6LoW-PAN which leverages standard protocols to obtain a softwarized network. Specifically, our solution uses the mesh-under feature of 6LoWPAN as the forwarding mechanism to improve performance in case of fragmentation and, to the best of our knowledge, we are the first to adopt this mechanism in this context. We also use a subset of RPL functionalities for network setup and to provide basic connectivity with the SDN Controller. In particular, RPL is exploited only to build upwards routes, i.e. from nodes to the root, whereas downward routes are managed by means of the SDN Controller. This also solves most of the problems raised by exploiting RPL for M2M communications, as outlined in [1]. Finally, we design an OpenFlow-like configuration interface that exploits the Constrained Application Protocol (CoAP) [9] for message exchange between the SDN Controller and the SDN

Nodes. To demonstrate the feasibility of our approach, we validate our system first on an emulated environment, i.e. by exploiting Cooja, then by using a real implementation on a testbed. Our results show that the additional communication costs due to control-plane messages is limited, and does not impair significantly the SDN solution in real scenarios. Moreover, we show that in case of M2M the SDN solution outperforms the standard RPL one in terms of communication delays.

The rest of the paper is organized as follows: in Section II we describe the technology background of WSN. In Section III we discuss the design ideas behind our solution, whereas in Section IV and V we detail its architecture and Control Plane protocol, i.e. the Southbound-Interface. In Section VI we describe how we validate our system via both simulation and prototyping. In Section VII we review the related works on the topic. Finally, Section VIII draws the conclusion and closes the paper.

## II. BACKGROUND

In this section we introduce the main protocols that are used in our solution, focusing on the unique characteristics exploited in our solution.

### A. 6LoWPAN

6LoWPAN has been developed by IETF in order to enable IPv6 end-to-end communication on top of 802.15.4 networks and it works as an adaptation layer between the IPv6 layer and the 802.15.4 MAC layer. Specifically, the 6LoWPAN layer adapts IPv6 packets to 802.15.4 frames by employing different techniques ranging from header compression by means of cross-layer optimizations, i.e. it uses information in the link and adaptation layers to compress network- and transport-layer headers, to packet fragmentation. Indeed, the IPv6 standard does not allow packet fragmentation and it requires an underlying channel able to transmit packets of at least 1280 byte, while the maximum packet size for 802.15.4 is 127 bytes. 6LoWPAN solves this issue by introducing Furthermore, IPv6 has been designed around the concept that the link is a single broadcast domain, such assumption is not valid for WSNs usually characterized by mesh of short-range connections.

To address these issues the 6LoWPAN layer introduces three main functionalities: i) Header compression, ii) Packet fragmentation and iii) Layer-two forwarding. Specifically, the Header compression is exploited to compress or elide IPv6 header fields that can be derived from link-level information carried in the 802.15.4 frame. Fragmentation is used to split IPv6 packets into multiple link-level frames to accommodate the IPv6 minimum MTU requirement. Finally, by means of the Layer-two forwarding, the adaptation layer can carry link-level addresses for the ends of an IP hop in order to expose a single point-to-point link to the IP layer regardless of the final IP destination. All the three techniques can be combined by means of stackable headers that are added by the 6LoWPAN layer to the 802.15.4 frames. As an example, the Mesh Header is exploited for Level-two forwarding and it is composed by two additional addresses, *final-address* and *originator-address*, that indicate the end-points of an IP hop.

### B. RPL Routing Protocol

RPL is an IPv6-based routing protocol specifically designed for lossy environments and resource-constrained embedded devices. Specifically, RPL employs a distance vector routing algorithm which builds a logical topology on top of the physical network. In particular, the topology is a Destination Oriented Directed Acyclic Graph, DODAG for short. The root node of the DODAG initializes the DODAG formation by emitting DODAG Information Object messages (DIO messages). Non-root nodes listen for DIOs and use the included information to join a DODAG. As a node joins a DODAG, it starts advertising its presence through the emission of DIO messages. Each DIO message specifies the rank of the sender, which is a scalar measure of the distance of that node from the root. As DIO messages are received from the neighbors each node updates its view of the topology and selects a preferred parent, used to forward traffic towards the root, among the set of neighbors with lower rank. DIO messages are, therefore, exploited to build upward routes.

On the other hand, downward routes are built by means of Destination Advertisement Object (DAO) messages, which are periodically propagated upward (along the DODAG) and are exploited to announce prefixes reachable by each node. Specifically, two different mode of operation are allowed: *storing* and *non-storing*. In the first mode, each node exploits received DAO message to build a routing table with all the routes reachable by means of its sub-DODAG. In non-storing mode, DAO messages are always forwarded up to the root that is the only network node with a routing table. In such mode routing information are included within each packet. For M2M communication the string mode of operation is therefore preferable because packets do not need to traverse the overall network up to the DODAG root, however, as reported in [1], the storing mode intrinsically introduces serious scalability issues, in particular for nodes close to the root, which must maintain potentially huge routing tables. Furthermore, in [1], it is also reported that point-to-point communications, i.e. M2M communications, suffer from poor reliability and such poor performance may prevent future adoption of RPL in the ever-increasing IoT applications scenario.

### C. Constrained Application Protocol

The Constrained Application Protocol (CoAP) is a lightweight RESTful protocol designed by the IETF as the reference application protocol for constrained devices. It is based on UDP, it inherits the same client/server paradigm adopted in HTTP, but it exploits a binary encoding of messages in order to reduce the communication overhead. The protocol is internally composed by two sub-layers: a request/response sub-layer and a message sub-layer. The former sublayer handles CoAP requests by invoking application methods to generate CoAP responses. As a RESTful protocol CoAP exposes a CRUD interface composed by four different methods: GET, POST, PUT and DELETE. The message sub-layer, instead, manages the exchange of messages between CoAP endpoints over UDP. To this aim, the sub-layer detects duplicates messages and manage the optional reliable delivery. The latter, in particular, is a simple stop-and-wait mechanism that add reliable transport on top of UDP by means of ACK messages and retransmissions.

III. DESIGN OF SD-6LOWPAN

In this section, we describe the principles that guided the design of the SD-6LoWPAN system, addressing three main problems, namely the implementation of the southbound interface, the management of the network-formation and the realization of an efficient forwarding mechanism. As we highlighted in the introduction, our high-level goal is to develop an efficient system based on standard protocols only, to ease the development of a real testbeds and to allow a smoother integration with existing systems.

*A. Control plane*

To manage control-plane operations in a centralized manner, every SDN Node must communicate with the remote SDN Controller in a bi-directional way. On one hand, the SDN Controller has to install the forwarding rules into SDN Nodes, on the other hand, SDN Nodes have to communicate their view of the network neighborhood to the SDN Controller, allowing the latter to build a *global* view of the network. The interface that implements this communication is called Southbound Interface (SBI) in SDN terminology. The de-facto standard communication protocol for SDN, OpenFlow, provides an SBI that is not suitable for the IoT environment [6], since it is designed for wired networks and it does not address common limitations of WSNs. Control and data messages must share the unique available communication channel. Therefore, efficient signaling protocols are needed in order to reduce the overhead introduced by control plane messages. In this work, we focus on exploiting standard protocols and solutions in order to simplify the integration between our proposal and currently available WSNs. For this reason, we design and implement the southbound interface following the principles defined by OpenFlow. Moreover, in order to reduce the overhead whilst maintaining a standard approach, we exploit the Constrained Application Protocol (CoAP) for control plane messages. In particular, thanks to the RESTful nature of CoAP, the SBI is designed by means of *resources* residing in both the SDN Controller and SDN Nodes. Each resource realizes a specific functional component of the SDN system (e.g. Flow Table, Neighbors Table and so on) and the IETF Concise Binary Object Representation (CBOR) [12] is exploited, whenever possible, to compress the payload of control messages. The use of a stateless client-server approach is in line with other widely used network-management protocols, such as SNMP [14], or the more recent NETCONF [16]. Moreover, the IETF has also standardized RESTCONF [15], which is a management protocol that exploits a RESTful approach based on HTTP, whereas the CoAP Management Interface (CoMI) [13], which extends RESTCONF and NETCONF over CoAP, is in the draft phase of the IETF standardization process. Finally, it should also be noted that CoAP implementations are available almost on any embedded OS. As an example, the Contiki OS provides both a server- and a client-side implementation of CoAP under the name Erbium. We will provide a detailed description of these resources in Section V.

*B. Topology Maintenance*

The SBI described in the previous section, allows SDN nodes to communicate their local view of the network to the SDN Controller. However, the SBI does not specify how to perform topology maintenance. The latter is a complex task that includes the following operations:

1. Network Discovery, i.e. discover the one-hop neighborhood of every SDN node and derive the overall network topology;

2. Network Formation, i.e. providing each SDN Node with a path towards the DODAG root, to allow basic connectivity at any given time towards the controller.

In OpenFlow, network discovery is done through the so called OpenFlow Discovery Protocol. The latter is not part of any OpenFlow standardization but is the de-facto standard for network discovery. Whenever a switch connects to a controller, the controller periodically commands the switch to transmit, over all its interfaces, discovery packets by means of the Link Layer Discovery Protocol (LLDP) and the Broadcast Domain Discovery Protocol (BDDP). Both discovery protocols are based on flooding mechanisms, thus introducing a significant communication overhead. Such approach is not feasible in WSNs where the flooding of packets must be reduced to the minimum. Therefore, our solution exploits the broadcast nature of the wireless medium to populate the internal neighbor table of each node: each node overhears incoming packets and creates a neighbor list, i.e. a list containing all nodes within its reception range. This list is then notified periodically or on demand to the SDN Controller through the SBI.

Concerning instead network formation, we use the *DODAG formation and maintenance* feature of RPL. Specifically, data paths between each node and the DODAG root, used to enable control message exchange with the SDN Controller, are established locally based on the RPL DODAG. A detailed description of this aspect is provided in Section IV.

*C. Forwarding*

To allow packet forwarding between multi-hop paths in 6LoWPANs two different mechanisms may be employed: i) route-over, or ii) mesh-under. The former mechanism delegates the forwarding decision to the IP layer that analyzes the destination IP address and selects the next-hop according to the information stored by the routing algorithm. Specifically, each node acts as an IP router and each link between nodes is seen as an IP hop. Due to this, if a packet is fragmented at the MAC layer, each node must receive all fragments and reconstruct the packet before forwarding, thus introducing additional delay in the forwarding process. The mesh-under mechanism, instead, operates below the IP layer and the forwarding is accomplished based on MAC addresses. At the IP layer, all the nodes within the 6LoWPAN are at one-hop distance, thus simplifying the overall network management. Fragments are treated independently, and each fragment can be forwarded without reassembling the overall packet at each hop. To this aim a specific header, called 6LoWPAN Mesh Header, specifies, per each fragment, the link-layer originator and the final destination – see [10] for a detailed comparison between route-over and mesh-under.

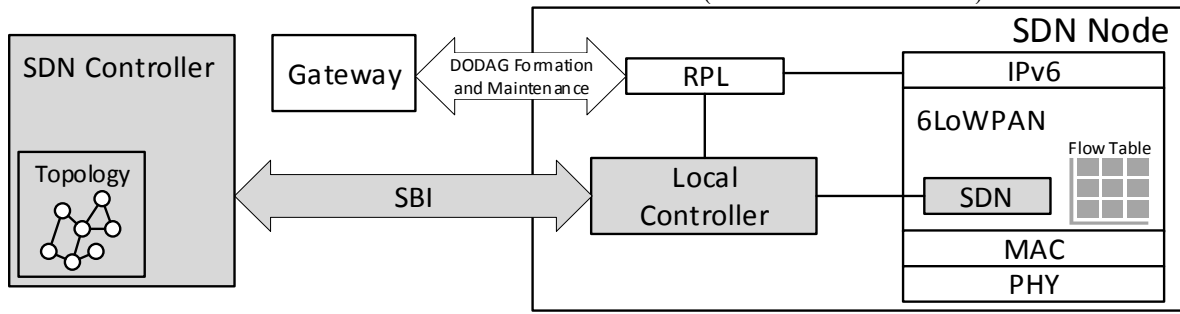

Fig. 1. System architecture of the SD-6LoWPAN. White blocks are standard layers, whereas grayed blocks one are custom ones.

Despite the introduced delay, most 6LoWPANs exploit a route-over solution to forward packets on multiple hops. One of the main reasons of such choice is the lack of a standardized layer 2 routing protocol which can be exploited by nodes to build the forwarding tables. In an SDN-based solution, however, the forwarding of packets is governed by a centralized SDN Controller that populates the forwarding table (flow table), on each node. Thanks to this the mesh-under solution can be employed in a seamless manner. Specifically, the SDN Controller retrieves all the information of the 6LoWPAN during the network formation procedure and fills the flow table of every node, by means of the SBI interface, with all the information needed to perform routing at layer 2, i.e. each packet is identified and forwarder according to the MAC addresses.

## IV. SYSTEM ARCHITECTURE

The system architecture is reported in Fig. 1 and is composed of three different main elements: i) the SDN Controller, ii) the Gateway, and iii) the SDN Nodes. The SDN Controller is exploited to configure and manage the control plane of the overall 6LoWPAN in a seamless and standard way. The SDN Nodes are standard sensor nodes enriched with an additional SDN layer. Finally, the Gateway is an SDN Node that also implements RPL border router capabilities. In the following we first describe these different components then we deeply describe the protocol stack.

*A. SDN Controller*

The SDN Controller manages the control-plane operations of the 6LoWPAN. It is physically outside the WSN but can access it through the Gateway. It gathers information on the network topology and on the quality of links among Sensor Nodes, e.g. the RSSI or ETX towards each neighbor, or regarding nodes itself, e.g. remaining energy or the average packet queue occupation. This information is used by the SDN Controller to generate a *global network view*. The latter is used to compute forwarding rules according to configurable policies, e.g. using shortest-path algorithms. Finally, the SDN Controller implements one end of the SBI, acting as a CoAP endpoint in the interaction with Sensors Nodes.

*B. Gateway*

The Gateway is an SDN Node that is used to provide connectivity between the 6LoWPAN and the rest of the network, which includes also the SDN Controller. As we described in the previous section, topology maintenance is provided via RPL. The Gateway therefore is configured as the *root* RPL and thus deployed as a DODAG root.

*C. SDN Nodes*

SDN Nodes are 6LoWPAN nodes enhanced with SDN capabilities provided by an additional sub layer. This layer provides network programmability by applying forwarding rules as specified in the so-called Flow Table, which is remotely managed by the SDN Controller. Every node runs an instance of the RPL protocol to discover the best route to communicate with SDN Controller. Therefore, the control plane is not totally separated from the data plane. A piece of control plane is placed at the SDN Node because of their distributed nature and impracticability in establish a separate channel between nodes and SDN Controller.

Protocol layers that compose the new architecture of the SDN-enabled sensor node are shown in Fig. 1. The additional sub-layer, called SDN, is placed within the 6LoWPAN adaptation layer. It handles data plane operations, i.e. packets forwarding according to the content of the Flow Table. The latter is instead populated by the Local Controller module through the SBI. It is worth noting that all forwarding operations are handled at the 6LoWPAN through the flow table, i.e. no IP routing is performed. The main tasks that each new layer must carry out are described in the following list.

*1) SDN Sub-Layer*

The SDN layer intercepts packets that flow between 6LoWPAN and MAC layers. As a packet arrives, it scans the Flow Table for a matching rule and forwards the packet accordingly. If no match is found, i.e. a *table-miss* has occurred, the SDN Layer notifies the Local Controller. It also defines conditions specifying if a traversing packet has to be handled by the Flow Table or not. E.g., packets that are addressed to the node itself and RPL messages are delivered to the 6LoWPAN layer without querying the flow table.

*2) Local Controller*

The Local Controller covers three main tasks: topology maintenance, topology update, SBI handling. The first task is to initialize and maintain connectivity with the on-hop neighborhood and with the SDN Controller, regardless of the availability of software defined routes. To this purpose it installs upstream rule in the Flow Table using the RPL protocol to obtain information regarding the preferred next-hop towards the Gateway. This allows nodes to know how to contact the remote SDN Controller in a distributed manner. It is worth to note that, using SDN, the 6LoWPAN is a *DAO-free* 6LoWPAN because only RPL DIO messages are needed to build upwards routes. Moreover, the Local Controller gathers information about neighbor nodes to install, for each of them, one-hop rules needed to forward packets addressed to direct neighbors.

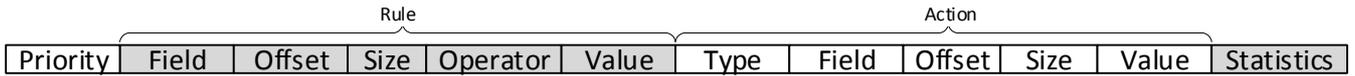

Fig. 2. Structure of a Flow-Table entry.

TABLE I. DESCRIPTION OF THE CTIONS AND RELATED FIELDS

| Type | Description | Field | Value | Offset | Size |
|---|---|---|---|---|---|
| *Forward* | Sends the packet towards specified next-hop. | - | MAC address of next hop. | - | MAC-address size in bit |
| *Broadcast* | Broadcasts the packet via layer-two broadcast frame. | - | - | - | - |
| *Modify* | Writes a value into the specified packet field. | Packet field to be modified | New value to written | Bit offset within the specified field | Number of bits to be written |
| *Drop* | Discards the current packet and prevents any other further actions on it. | - | - | - | - |
| *Decrement* | Decreases a packet field by certain value. | Packet field to be modified | Value of the decrement | - | - |
| *Increment* | Increases a packet field by certain value. | Packet field to be modified | Value of the increment | | |
| *To Upper Layer* | Delivers the packet to the upper layer. | - | - | - | - |
| *Continue* | Indicates that the matching procedure will continue after the current action is executed. | - | - | - | - |

The second task is to periodically report to the SDN Controller a local topology view containing a list of neighbor nodes with the respective link quality metric, and node state information such as battery level. This periodic operation is called *Topology Update*. In this manner, the Controller can obtain the global network topology and state information regarding each sensor node.

The third task is to manage the forwarding rules contained in the Flow Table, handling installation, removal, reading and modification of table entries. For this purpose, it implements the SBI through CoAP resources. It handles the interaction with the SDN Controller when a table-miss occurs by creating CoAP requests and parsing the corresponding response.

*3) Flow Table*

The Flow Table is the data structure that is used to associate data packets with forwarding decisions. It is organized in rows, each one comprising a matching rule and an action. Rules and actions are designed taking into account the data centric nature of WSNs, thus they consider not only protocol headers but also data payloads. The detailed structure of a Flow-Table entry is shown in Fig. 2 and is composed of four main sections: priority, rules, actions and statistics.

Priority is a number that determines the examination ordering of rows, as multiple ones can apply to one packet.

Rules are a list of conditions that must be satisfied by the packet to apply the associated actions. Rules within the same row are applied in *and* so all of them must be verified for the rule to apply. Each rule applies the *operator* to two operands: the first one is identified specifying the *field* within the packet, its starting bit *offset* and *size*, whereas the second one is specified by *value*. Offset and size are used to address a specific bit window within the overall field. The allowed operators are $=, \neq, \leq, \geq, <, >$. Entries without rules are allowed and they thus act as *default* entries.

The action section specified the sequence of operations that will be performed on the packet upon matching of all the corresponding rules list. If multiple actions are specified, they are executed sequentially. An action is described through 5 fields, wherein the *type* specifies the kind of operation to perform, and the meaning of the other fields changes accordingly. In TABLE I we summarize the main actions and associated meaning for the other relevant fields.

Finally, the Statistic section contains information on the usage of the corresponding entry. This section is updated every time the entry is accessed, i.e. when its rules are satisfied. It contains a *counter* that is incremented at each access, and a Time to Live that decreases over time and represents the remaining time before the entry is removed.

*D. Forwarding example*

In Fig. 3, we report an example of the forwarding mechanism employed in our solution. Assume that the node with ID 3 (FE80::3) sends a packet to the node with ID 7 (FE80::7). By employing a mesh-under solution the IP layer is totally unaware of the multi-hop path and creates the packet by filling the source and destination IPv6 addresses. The 6LoWPAN layer (Step 1) compresses the headers and adds the mesh header specifying the link-layer originator and the final destination. The SDN layer (Step 2) scans the flow table and finds the packet matches with a specific rule, which selects the MAC address of the node with ID 5 as the next hop. The MAC layer (Step 3) forwards the packet, with the additional headers, to node with ID 5 that receives the message and processes it in its SDN layer (Steps 4 and 5). The packet is then forwarded to node with ID 7 (Step 6) according to the information stored in Node 5's flow table. It is worth to note that the node with ID 5 processes the packets below the IP Layer, such behavior is allowed by the mesh-under forwarding mechanism and it introduces several advantages that cannot be obtained in a route-over solution. Indeed, in a route-over solution the rules inside the flow table must handle IPv6 addresses, therefore a packet must be completely reassembled and decompressed by the 6LoWPAN layer on each hop. Specifically, the 6LoWPAN layer must reassemble and decompress the packet in order to pass it to the IP layer, then, if the packet must be forwarded (as in case of node with ID 5), the 6LoWPAN layer must compress and fragment the packet again before forwarding it to the next hop.

## V. A RESTFUL-BASED SOUTHBOUND INTERFACE

As we anticipated in the previous sections, we implement an OpenFlow-inspired SBI based on a standard communication protocol, i.e. CoAP. The nodes involved in the SBI, namely SDN Controller and SDN Node, expose a set of CoAP resources that together compose the SBI.

*A. SND-Controller Resources*

The SND controller exposes two resources: Network and Flow Engine. The former is the current state of the 6LoWPAN network including the set of nodes belonging to it, their attributes and how they are connected. This resource allows only POST requests. Every sensor node that implements the SDN architecture performs periodically a POST request on Network resource, to send information on its state and a list of reachable neighbor nodes, through Topology Update. When sent during the configuration phase, the response message body contains the configuration parameters for the controlled node. During nominal operations instead, it will be empty.

The Flow Engine resource handles the distribution of forwarding rules. It can be accessed via a GET or a POST operation. In the first case, the requester specifies the source's MAC address, and the resource replies with all entries associated to that node. This allows nodes to act in a proactive manner, obtaining rules previously defined for them, without waiting for a table miss to occur. When instead a table miss occurs, the Flow Engine resource is accessed via a POST operation, which includes the whole packet that triggered the table miss. To limit communication overhead, instead of sending the whole packet, the SDN Node can be configured to include only a set of packet's *key-features* in the POST method, which in turn are specified via a dedicated CoAP resource of the SDN Node.

*B. SND-Node Resources*

The SDN Node exposes a set of CoAP resources that allow the SDN Controller to configure how the former uses the SBI and to query the former's neighbor table. There are a total of five CoAP resources available.

The *Flow Table* resource encapsulates the Flow Table data structure lets the SDN Controller to retrieve and modify the current table content. It allows three operations: a GET one returns the entire set of Flow Entries contained in the table. The table rows are formatted in the response using a CBOR to reduce overhead. A PUT operation installs new rules into the Flow Table. Finally, a DELETE operation removes *all* the rules contained in the forwarding table.

The *Update-Period* resource regulates the period of the Topology Updates. It can be either read through a GET request or set to a new value through a POST one.

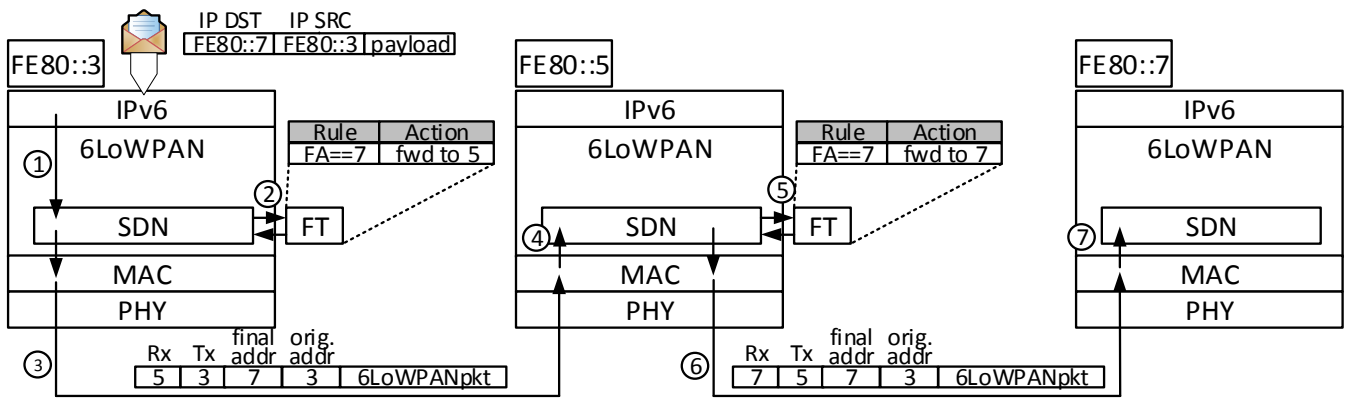

Fig. 3. Example of forwarding based on SD-6LoWPAN.

The *Key-Feature* resource specifies which fields of the packet that caused a table miss have to be sent to the SDN Controller. This resource allows one to reduce the network overhead during table misses, as it avoids transmitting the whole packet to the SDN Controller. This resource can be read and set through a GET and POST request respectively.

Finally, the *Neighbors-Table* resource exposes the table of neighboring table of the SDN node, having for each entry the MAC address and statistics on wireless link quality as the RSSI and ETX. This resource allows the SDN Controller to obtain information regarding the network topology without waiting for the Topology Update message. The Neighbors-Table resource can be accessed only through GET requests.

## VI. VALIDATION AND PERFORMANCE EVALUATION

As a first step, we validate our solution in a simulated environment to assess the feasibility of the proposed SDN architecture. To this aim, we exploit the Cooja simulation environment and we design a set of experiment to compare our SDN-based solution against a standard RPL network. Then, we evaluate the proposed solution in a real testbed deployed at our department.

### A. Cooja simulations

This first experiment has been designed to compare our SDN solution against a standard solution that exploits IP forwarding and RPL. We run the two configurations on the network in Fig. 4 to evaluate the SDN overhead, in terms of the number of transmitted control messages. We use the Unit Disk Graph Medium (UDGM) channel model and the Cooja motes. To achieve statistical soundness, we run 20 independent replicas, each one one-hour long. The SDN controller implements the Dijkstra algorithm, using ETX as metric, to populate the flow tables of SDN nodes.

As can be seen in Fig. 4, there are three different types of nodes: i) sending nodes (white), ii) forwarder nodes (grayed), and iii) RPL border router (dotted). Each sending node periodically sends an UDP message to the external UDP Server which replies with the same message. UDP messages have a fixed payload of 40 bytes, and are transmitted periodically having a period randomly selected between 30 and 90 seconds. Forwarder nodes, instead, do not generate UDP traffic, but are used only to forward messages within the 6LoWPAN, thus acting as communication relays. Finally, the RPL border router connects the 6LoWPAN to the external network where both the UDP Server and the SDN controller are hosted. TABLE II reports a summary of the main simulation parameters. It is worth noting that, when SDN is enabled, DAO messages are useless and thus we removed them to have a DAO-free 6LoWPAN.

To analyze the obtained results, we first compute the time needed by the network to converge to a *steady state*, i.e. all flow tables are populated by the SDN Controller. Indeed, at the beginning flow tables are empty and each incoming packet generates a table-miss, which is notified to the SDN controller. According to our results we discovered that after about 15 minutes the network reaches stability, therefore in the following we present only the results collected during the *steady state* in order to provide a fair comparison between SDN and RPL solutions.

In Fig. 5 we report the overhead due to control messages counting for, in the RPL case, the RPL DIO and RPL DAO messages, whereas, in the SDN case, the RPL DIO and SDN messages, i.e. topology update messages and table miss messages. As can be seen the SDN solution always introduces additional communication costs mainly due to the topology update messages and table misses periodically generated by changes in the network topology. However, this overhead has a small impact on data packets as we show in Fig. 6, where we report the mean round trip time (RTT) experienced by UDP packets. As can be noted, RPL and SDN achieve similar results and on average the SDN solution slightly outperforms the RPL one. This can be explained considering the different mechanism employed for forwarding packets. Specifically, in the RPL case each node exploits the standard route-over solution that requires to process each packet at the IP layer. In the SDN case, instead, the forwarding of packets is accomplished by means of the flow table below the 6LoWPAN layer, thus removing the time needed to compress/decompress packet headers.

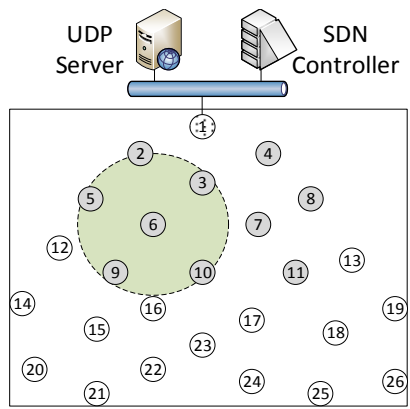

Fig. 4. Network topology used in the Cooja simulations

TABLE II.     MAIN SIMULATION PARAMETERS

| SDN | |
|---|---|
| Flow Table size | 40 Flow Entries |
| Topology Update period | 20 min |
| DAO messages | Disabled |
| **RPL** | |
| IPv6 Routing Table size | 40 Routes |
| DAO messages | Enabled |

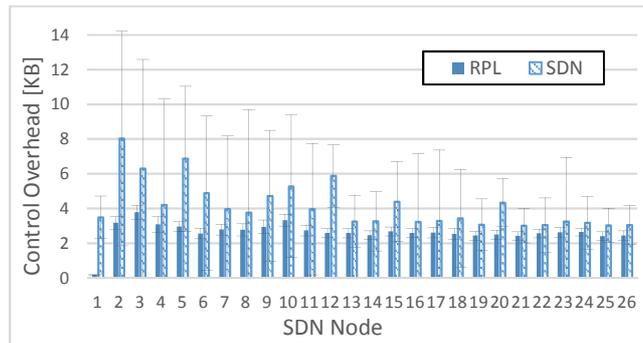

Fig. 5. Comparison of the control-messages overhead in the SDN and RPL scenarios, obtained with Cooja simulations.

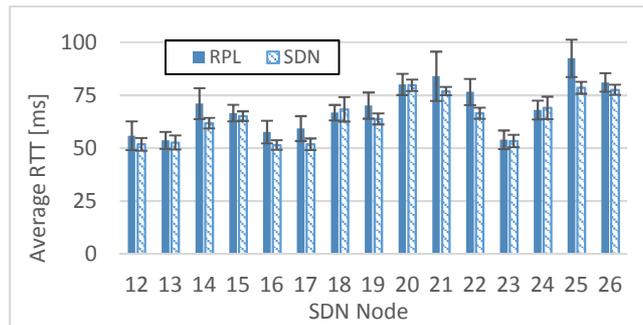

Fig. 6. Comparison of the RTT experienced by data packets in the SDN and RPL scenarios, obtained with Cooja simulations.

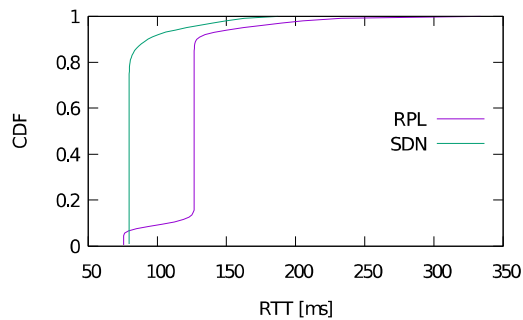

Fig. 7: CDF of the RTT experienced by M2M data packets in the SDN and RPL scenarios, obtained with Cooja simulations.

However, employing SDN can effectively reduce the RTT in M2M communications compared to the standard RPL case. To highlight this, we perform a slightly different experiment based on the same network topology, configuration and traffic pattern of Fig. 4 and TABLE II. The only difference is that node 20 generates periodic traffic towards node 26 in which we deploy an UDP Server application that replies exactly in the same manner as the external UDP Server of Fig. 4. The other sender nodes continue to generate periodic traffic for the external UDP Server. We repeat the experiment 15 times.

In Fig. 7, we report the empirical cumulative distribution function (ECDF) of the RTT experienced by node 20 to communicate with node 26. Due to the simulation environment and the fact that all communications are within the WSN, the delay is defined almost only by the number of hops. Nevertheless, thanks to the SDN Controller which setups a direct path between the two involved nodes, the experienced RTT does not change between replicas and the 83% of packets experience a delay below 80 ms. The RPL case, instead, is much more affected by different replicas, showing a bimodal distribution of the CDF. Specifically, we setup RPL in s*toring* mode, thus each M2M data packet starts to traverse the DODAG up, until a common ancestor between node 20 and node 26 is found. The packet is then forwarded down until it reaches node 26, which replies back. Due to this, depending on how the DODAG is formed the number of hops changes between replicas and therefore the experienced delay. Nevertheless, as can be seen, more than 90% of packets are received with a delay greater than 120 ms, whereas only 10% of packets experienced a delay comparable to the SDN case, as in few replicas the path selected by SDN and RPL are composed by the same number of hops.

TABLE III. SDN SOLUTION COMPARISON

|  | SDN-WISE [7] | SD6WSN [11] | µSDN [8] | SD-6LoWPAN |
|---|---|---|---|---|
| Control Plane | Ad-hoc | CoAP | UDP-JSON | CoAP |
| Forwarding | Ad-hoc | Route-over | Route-over | Mesh-under |
| Routing | Ad-hoc | RPL | RPL | RPL |

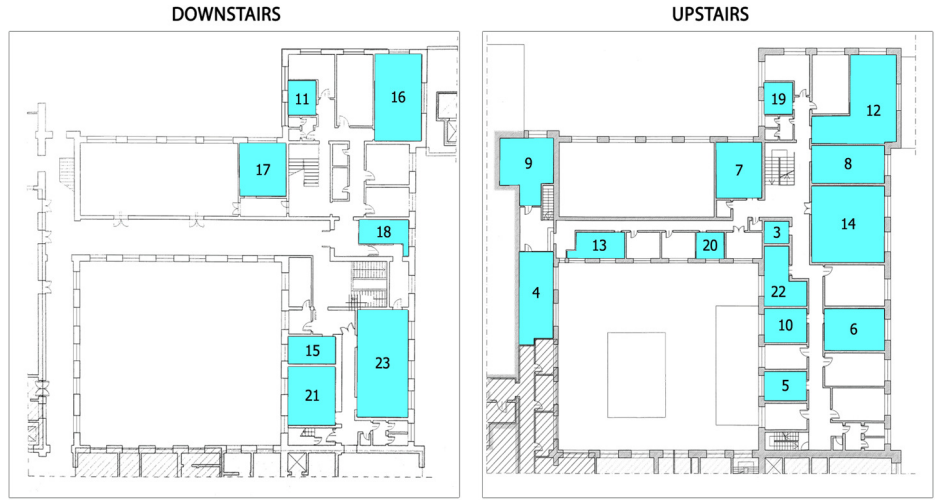

Fig. 8. Network deployment of the the testbed.

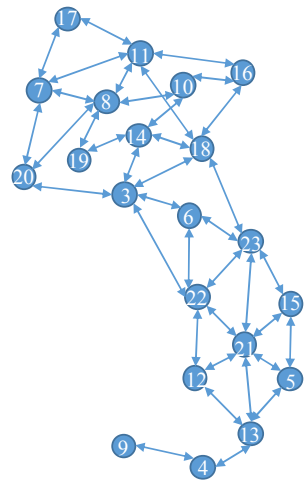

Fig. 9. Logical topology resulting from the testbed deployment.

*B. Testbed implementation*

We now evaluate our solution on real hardware via the IoT testbed deployed within Department of Information Engineering of the University of Pisa. The testbed network is composed of 21 RE-Mote sensor nodes, each one equipped with an ARM Cortex-M3, 512 KB of flash and 32KB of RAM. A detailed description of the testbed is available in [17]. Nodes are placed in multiple rooms over two floors. The nodes deployment and the derived network topology are reported in Fig. 8 and Fig. 9 respectively.

Compared to the simulated scenario, we modified a subset of the configuration parameters to overcome the limited resources on sensor nodes. More in details, the Flow Table size and the Routing table size have been reduced to 20 entries, whereas the Topology update period has been set to 10 minutes in order to better handle dynamic changes in the network topology due to interferences. For the same reason, we replaced the hop-count policy in the SDN controller with a more enhanced policy that takes into account the measured ETX between nodes. Finally, we assume that each Node acts as a CoAP client that periodically – we kept the same period as the simulated scenario - issues POST messages to a CoAP server hosted on the same network of the SDN controller.

In Fig. 10 we report the overhead introduced by the SDN solution compared to the RPL case by measuring the number of bytes exchanged for control messages. As in the simulated environment, we report the results collected in the *steady state*. Even in this case the SDN solution introduces a sensible overhead compared to the RPL scenario, which is further increased by the smaller topology update interval. Moreover, the mean RTT increases as can be seen in Fig. 11. This can be explained considering that real sensor nodes are constrained devices with limited computational capabilities, thus control plane signaling can effectively introduce additional delay also for data packet handling.

## VII. RELATED WORK

In this Section, we analyze works available in the literature that integrate SDN in WSN highlighting the design ideas and the architectural solutions that are proposed. A first effort in the integration of SDN has been carried out in [6], where the authors propose an OpenFlow-like solution called Sensor OpenFlow. The main idea was to exploit the OpenFlow Extensible Matches to define two new types of matching classes specifically designed for WSNs. The proposed work, however, does not provide a working implementation on top of simulated or physical sensor nodes. Moreover, IP-connectivity is addressed only marginally.

Similarly, SDN-WISE [7], is a complete SDN solution for WSNs that totally redefine the traditional network stack and runs on top of the IEEE 802.15.4 MAC layer. A working prototype in Contiki OS is available. Thanks to the cross-layer approach of SDN-WISE radio duty-cycle and data-aggregation are natively supported and can be managed through the Controller. Whenever a packet is processed at a node, an internal WISE Flow Table is used to select the next hop and in case of table miss the Controller is queried. To this aim, each node maintains a path to the sink node. Such path is built by means of an ad-hoc protocol, called Topology Discovery (TD), that relies on the periodic exchange of packets containing state information (e.g. battery level and hop-distance from the sink). SDN-WISE is a complete solution, however, by exploiting only custom protocols it cannot be easily integrated within other solutions and an application gateway is always needed to translate packets between the WSN and external applications.

A further step has been achieved by the authors of [11] wherein the SDN approach is integrated into 6LoWPAN networks. The solution, called SD6WSN, leverages on the standard IETF stack composed by IPv6, RPL, UDP and CoAP. Specifically, each node implements: i) an Agent to communicate with the external controller, ii) a Data Plane Forwarding to store the flow table, and iii) a Control Plane Routing to maintain the path between the node and the RPL border router. The interaction with the external controller is handled by an ad-hoc protocol, called SD6WSNP, which is transported by means of CoAP messages. Packet forwarding, in particular, leverages route-over but, differently from other works that exploits source routing, the authors decided to exploit an ad-hoc solution to intercept UDP packets traversing the network. This design choice, however, implies a modification to the standard protocol layer in a cross-layer fashion and it cannot be easily extended to non UDP messages.

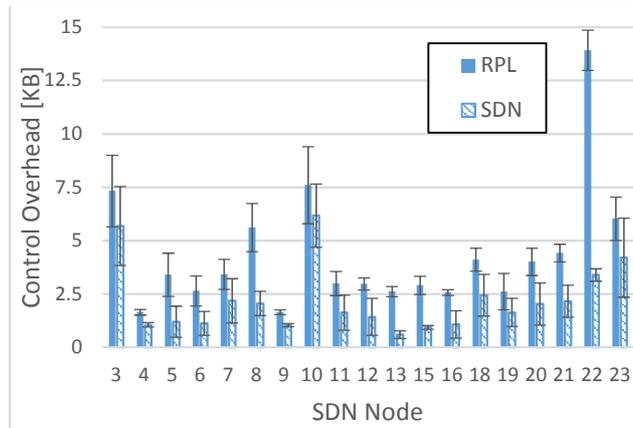

Fig. 10. Comparison of the control-messages overhead in the SDN and RPL scenarios, obtained with the testbed.

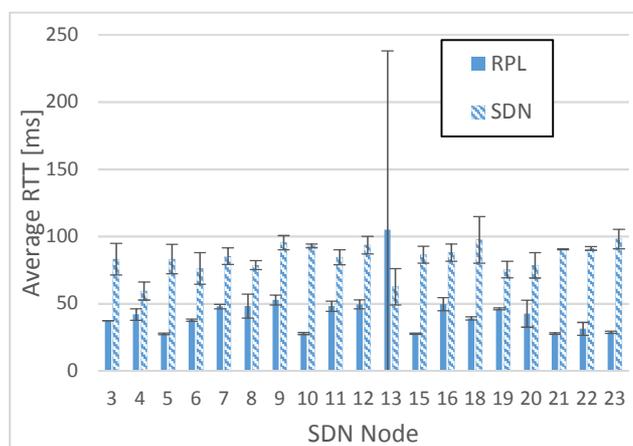

Fig. 11. Comparison of the RTT experienced by data packets in the SDN and RPL scenarios, obtained with the testbed.

Finally, another IETF compliant solution, called μSDN, has been presented in [8]. The authors define a dedicated application protocol, to handle communications between the node and the controller (implemented on the sink node of the WSN), as well as a new protocol layer that is introduced above the IPv6 layer. Such new layer implements the core logic and hosts the flow table populated by the controller. The adopted solution employs route-over on top of the RPL routing protocol which is employed to communicate, by means of DAO messages, to the controller the WSN topology and also to provide a default routing mechanism to reach the controller. Specifically, each sending node adopts a source-routing strategy which defines all the nodes that must be traversed by each packet. It is worth to note that, several ad-hoc optimizations have been carried out in order to reduce the size of messages within the WSN. All of them, however, have been defined to handle the specific use case and does not rely on standard protocols employed, nowadays, in the IoT field.

In TABLE III we report a comparison of the above mentioned SDN solutions. It is worth to note that our SD-6LoWPAN solution is the first solution that employs a mesh-under technique for data packet forwarding, employs only standard protocols and does not require any cross-layer mechanism.

## VIII. CONCLUSIONS

In this work we described SD-6LoWPAN, a system architecture for a smooth integration of SDN into 6LoWPANs. Our solution takes advantages of the existing technologies and protocols available in the 6LoWPAN context to create an efficient and flexible communication system, that can be deployed in real environments and can operate in conjunction with existing deployments. We validate our system through both simulation in Cooja, and via evaluation on a real testbed at our department. Results show that, the SDN solution introduces additional communication overhead due to the control message exchange with the controller but this does not increase communication delay significantly. Moreover, for M2M communications, communication delays can be instead significantly reduced, thanks to the establishment of direct routes between communication endpoints.

As a future work we aim at extending our implementation to reduce the SDN overhead, e.g. by exploiting Partial Packet Queries (PPQ) solutions in order to reduce the size of messages that are forwarded to the controller whenever a table miss occur. Moreover, we plan to extend the performance evaluation, including a thorough comparison with other solutions available in the literature.

## ACKNOWLEDGMENT

Work partially supported by the Italian Ministry of Education and Research (MIUR) in the framework of the CrossLab project (Departments of Excellence).

Authors would like to thank Giulio Micheloni and Simone Tavoletta, who contributed to the deployment of the SD-6LoWPAN system.